\documentclass[journal=jacsat,manuscript=article]{achemso}

\usepackage{chemformula} 
\usepackage[T1]{fontenc} 
\usepackage{array}
\usepackage{amssymb}
\usepackage{hyperref}                                   

\author{Shiang-Yu Huang}
\affiliation{Institute for Functional Matter and Quantum Technologies, University of Stuttgart, 70569 Stuttgart, Germany}
\email{shiang-yu.huang@fmq.uni-stuttgart.de}
\author{Stefanie Barz}
\affiliation{Institute for Functional Matter and Quantum Technologies, University of Stuttgart, 70569 Stuttgart, Germany}
\alsoaffiliation{Center for Integrated Quantum Science and Technology (IQST), University of Stuttgart, 70569 Stuttgart, Germany}
\email{stefanie.barz@fmq.uni-stuttgart.de}

\title[An \textsf{achemso} demo]
  {\vspace{-1.0cm}Compact Inversely-Designed Vertical Coupler with Bottom Reflector for Sub-Decibel Fiber-to-Chip Coupling on Silicon-on-Insulator Platform}

\begin{document}

\begin{abstract}
  Inverse design via topology optimization has led to innovations in integrated photonics and offers a promising way for designing high-efficiency on-chip couplers with a minimal footprint.
	In this work, we exploit topology optimization to design a compact vertical coupler incorporating a bottom reflector, which achieves sub-decibel coupling efficiency on the 220-nm silicon-on-insulator platform.
	The final design of the vertical coupler yields a predicted coupling efficiency of -0.35 dB at the wavelength of 1550~nm with a footprint of 14~\textmu m $\times$ 14~\textmu m, which is considerably smaller than conventional grating couplers.
	Its topology-optimized geometry can be realized by applying one full etch and one 70-nm shallow-etch process and the fabricability is also guaranteed by a minimum feature size around 150~nm.
	The proposed vertical coupler can further miniaturize photonic integrated circuits and enable highly-efficient networks between optical fibers and other photonic devices.	
\end{abstract}

\section{Introduction}

Photonic integrated circuits have been a promising platform for practical implementation in various fields,
especially for emerging quantum technologies such as quantum key distribution \cite{sibson2017chip,bunandar2018metropolitan,zhang2019integrated,beutel2021detector, zhu2022experimental,beutel2022fully,sax2023high},
quantum communication \cite{paraiso2021photonic,trenti2022chip} and photonic quantum computing \cite{qiang2018large,arrazola2021quantum,vigliar2021error,qiang2021implementing,chi2022programmable},
due to their scalability and mechanical and thermal stability.
Since the monolithic integration of the essential building blocks of a photonic quantum system, namely single photon sources, photonic components that process quantum information and single-photon detectors, is still technically challenging,
on-chip couplers, particularly the ones with sub-decibel coupling efficiency in telecom C band, 
are pivotal in interfacing the integrated quantum photonic circuits and optical fibers.
They enable efficient networks to other photonic integrated systems or off-chip devices without losing a great amount of single photons.

On-chip couplers can be generally divided into two categories:
in-plane couplers \cite{hatori2014hybrid,he2020low,mu2020edge} and out-of-plane couplers \cite{zaoui2014bridging,marchetti2017high,hoppe2019ultra,marchetti2019coupling,cheng2020grating}.
Compared to in-plane couplers like edge couplers, out-of-plane couplers or grating couplers offer relatively high alignment tolerance, enable wafer-scale inspection or testing and can be placed flexibly on the chip, easing the complexity of the circuit design.
They neither require special treatment, e.g.~cleaving and polishing processes, at the edge of the chip to create a quality facet, which lowers the difficulty in chip fabrication.
Despite the advantages mentioned above, conventional grating couplers usually require a non-zero fiber coupling angle to mitigate the loss caused by the back reflection from the second-order diffraction \cite{taillaert2006grating}.
Such a coupling scheme is not only disadvantageous in alignment routines but also introduces difficulties and additional costs to the packaging of the photonic integrated circuits.
Although conventional grating couplers using the vertical coupling scheme, where the fiber coupling angle is zero, have been demonstrated \cite{watanabe2017perpendicular,zhang2019high,liu2022high,xiong2022compact}, they usually require a long tapered waveguide, typically in the length of 40 µm to 200 µm, to reduce the loss during mode conversion into the waveguide fundamental mode.
Such a characteristic inevitably hinders the miniaturization of photonic integrated circuits for high-density integration and scalability. 

In this context, inverse design via topology optimization becomes a favorable method as it promises compact integrated components with comparable or even better performance compared to the traditional counterparts.
Generally, this technique varies the spatial distribution of the refractive index to optimize the user-defined figure-of-merit (FoM) functions that relate to the device performance.
Specific requirements, such as central operating wavelength, thickness of the cladding and device layers, etc., can also be considered in advance to simplify the entire design workflow.
Numerous novel designs of photonic integrated components, such as wavelength-dependent demultiplexers \cite{vercruysse2019analytical,piggott2020inverse}, power splitters \cite{piggott2020inverse,hansen2024inverse}, spatial mode multiplexers \cite{piggott2020inverse,shang2023inverse} and grating couplers \cite{dory2019inverse,hammond2022multi,pita2022ultracompact,wang2024single}, have been investigated or experimentally demonstrated on a variety of material platforms like diamond and lithium niobate.
Among these platforms, silicon-on-insulator (SOI) is an attractive one for scalable implementation since it leverages the mature complementary metal-oxide semiconductor (CMOS) manufacturing technologies.
Topology-optimized SOI couplers using the vertical coupling scheme have also been simulated and experimentally demonstrated~\cite{hammond2022multi,pita2022ultracompact,wang2024single}.
Despite their compact footprint, however, the designs presented in Refs. [\!\!\citenum{hammond2022multi,pita2022ultracompact,wang2024single}] exhibit a relatively high simulated coupling loss of -3.0 dB, -1.9 dB and -1.1 dB at the wavelength of 1550~nm, respectively.
This issue presumably arises from the high substrate leakage \cite{hammond2022multi,pita2022ultracompact} or the strong back reflection due to the fully-etched grating geometry \cite{wang2024single}.
To facilitate ultrahigh-efficiency vertical fiber-to-chip coupling with a minimal spatial footprint, a new design that can address the aforementioned issue is imperative.

In this work, we apply the topology optimization to inversely design a 14 µm $\times$ 14 µm vertical coupler which has a shallow-etched geometry on the 220-nm SOI platform (Fig.~\ref{fig:sim_sketch}(a)).
The topology-optimized vertical coupler incorporates a bottom reflector to improve the diffraction directionality.
The simulated coupling efficiency of the acquired design is -0.35 dB (92.2\%) at 1550~nm and also has a flat spectral response within the telecom C band.
Compared to the conventional vertical grating couplers in Refs. [\!\!\citenum{watanabe2017perpendicular}] and [\!\!\citenum{liu2022high}] with the state-of-the-art coupling efficiency of -0.6 dB and -0.75 dB in simulations, respectively, our design is predicted to have improved performance with a significantly reduced spatial footprint.
Further details of the aforementioned SOI vertical couplers can be found in the comparison shown in Table~\ref{table: coupler comparison}.
We anticipate our new design will further extend the applications of the topology-optimized couplers in various photonic applications and chip networks requiring high-efficiency vertical fiber-to-chip coupling.

\renewcommand{\arraystretch}{1.5}
\begin{table}[h]
    \centering
    \scalebox{0.83}{

    \begin{tabular}{ || >{\centering\arraybackslash}m{4em} >{\centering\arraybackslash}m{4.5em} >{\centering\arraybackslash}m{12em} >{\centering\arraybackslash}m{3.5em}  >{\centering\arraybackslash}m{4em} >{\centering\arraybackslash}m{1.5em} >{\centering\arraybackslash}m{6em} >{\centering\arraybackslash}m{5em}||}
        \hline
    Year & Ref. 								 & Structure features 									     & Bottom reflector & Sim. CE \newline (dB) & Etch step &  Device layer thickness (nm) & Dimension \newline (\textmu m$^2$)\\  [1ex]
    \hline\hline
    
    2017 & [\!\!\citenum{watanabe2017perpendicular}]& Blazed GC with taper waveguide		   & \texttimes & -0.6 					&	2	& 220 	& $\sim$200 $\times$ 14 \\ 
    \hline
        2019 & [\!\!\citenum{zhang2019high}]& Bidirectional apodized GC with two taper waveguides		   & \texttimes & -1.42  					&	1 	 & 340	&  800 $\times$ 12   \\
    \hline
    2022 & [\!\!\citenum{liu2022high}]				 & Shallow-etched GC with taper waveguide   & \checkmark & -0.75 				&	2	& 220 	& $\sim$200 $\times$ 14 \\
    \hline
        2022 & [\!\!\citenum{xiong2022compact}]				 & L-shaped GC with in-plane metalens in taper waveguide   & \texttimes & -2.59 				&	2	 & 220	& 45 $\times$ 14 \\
    \hline
    2022 & [\!\!\citenum{hammond2022multi}]		 & Multilayer and topology-optimized  & \texttimes & -3.0 				 	&   3	& -	& 10  $\times$ 10   \\ 
    \hline
    2022 &  [\!\!\citenum{pita2022ultracompact}]	 & Shallow-etched and topology-optimized  & \texttimes & -1.9 					&	2	& 250	& 10  $\times$ 10  \\
        \hline
    2024 & [\!\!\citenum{wang2024single}]			 & Fully-etched and topology-optimized & \checkmark & -1.1 					&	1	& 220	& 2.8  $\times$ 2.8$^*$ \\
    \hline
        2024 & This work 							 & Shallow-etched and topology-optimized  & \checkmark & -0.35 				&	2	& 220	& 14   $\times$ 14  \\ [1ex] 
    \hline
    \end{tabular}
    }
    \caption{Merits of recent fiber-to-chip vertical couplers on the SOI platform. Sim.~CE: simulated coupling efficiency. 
  GC: grating coupler.
  $^{*}$Lens fibers with a mode field diameter of 2.5 \textmu m are applied.
  }
    \label{table: coupler comparison}
\end{table}

\section{Optimization and Simulation Results}
\begin{figure*}[h!]
  \centering
  \includegraphics[scale=0.36]{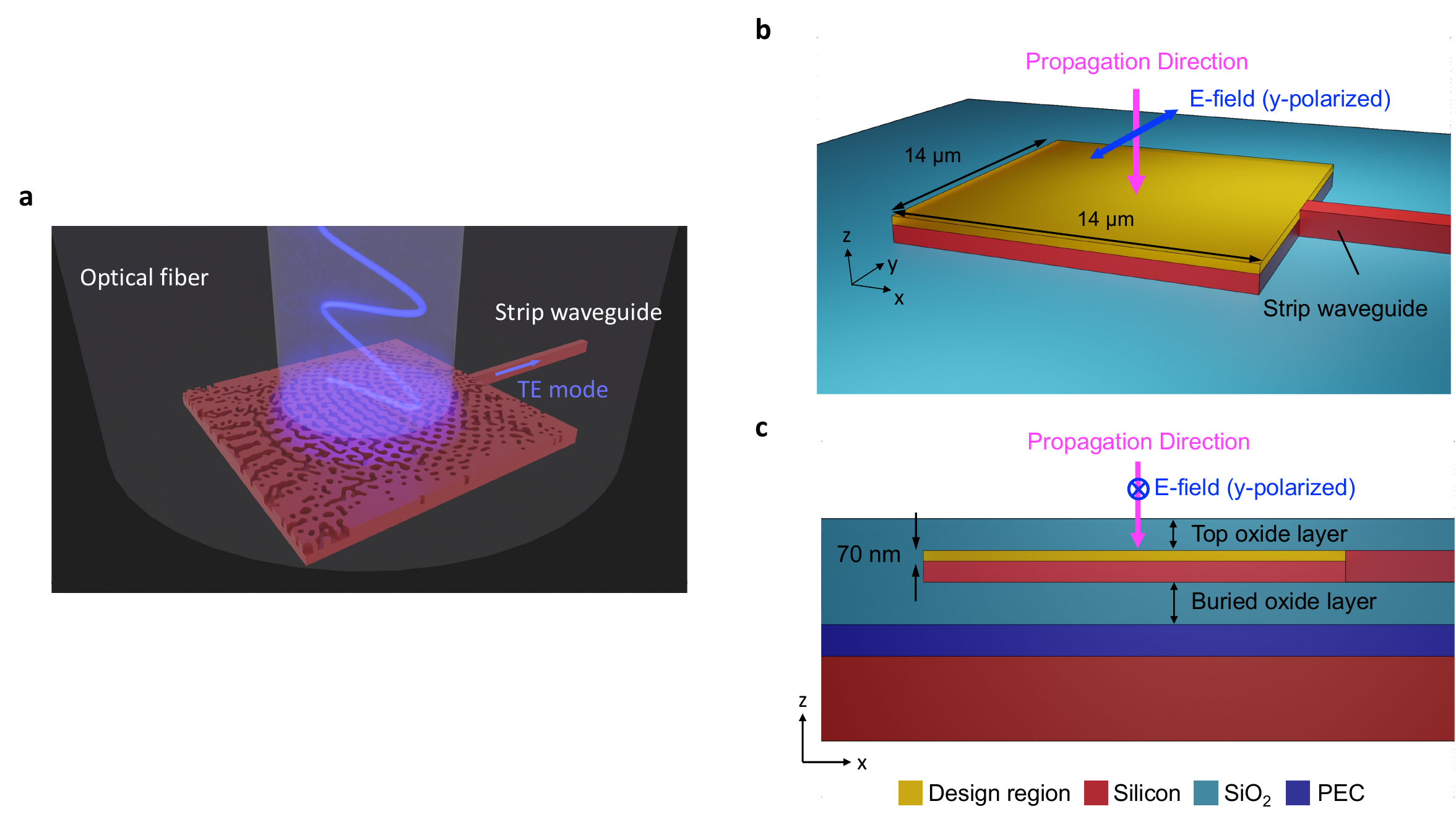}
  \caption{
    (a) Schematic illustrating the working principle and functionality of the topology-optimized vertical coupler.
  (b) Perspective view (omitting the top oxide layer) and 
    (c) the cross-section of the objects in the three-dimensional FDTD simulations carried out during the topology optimization (not to scale). 
  The size of the design region (yellow area) is 14 \textmu m $\times$ 14 \textmu m $\times$ 70 nm.
  The dimension of the strip waveguide is 450 nm $\times$ 220 nm on top of a 2-\textmu m buried oxide layer (cyan area below the silicon device layer).
  A layer of perfect electric conductor (PEC) is placed below the buried oxide layer and serves as a bottom reflector (blue area).
  }
  \label{fig:sim_sketch}
  \end{figure*}

We exploit the inverse design tool, Lumopt, together with the finite-difference time-domain (FDTD) solver (Ansys Lumerical FDTD) to perform the topology optimization and design the vertical coupler.
In a three-dimensional FDTD simulation environment, a design region with dimensions of 14 µm $\times$ 14 µm (Fig.~\ref{fig:sim_sketch}(b)) is set up to fully capture the incoming light from a standard single-mode optical fiber (e.g. SMF-28 with a mode field diameter of $\sim$ 10.4 µm at 1550 nm) and to obtain the maximal attainable coupling efficiency.
Furthermore, a perfect electric conductor (PEC) layer is inserted underneath a 2-\textmu m buried oxide layer (Fig.~\ref{fig:sim_sketch}(c)) and functions as a metal bottom reflector to enhance the directionality of the coupler.
Details of the settings regarding the simulation and topology optimization are illustrated in Methods.

\begin{figure}[hb!]
\centering
\includegraphics[scale=0.35]{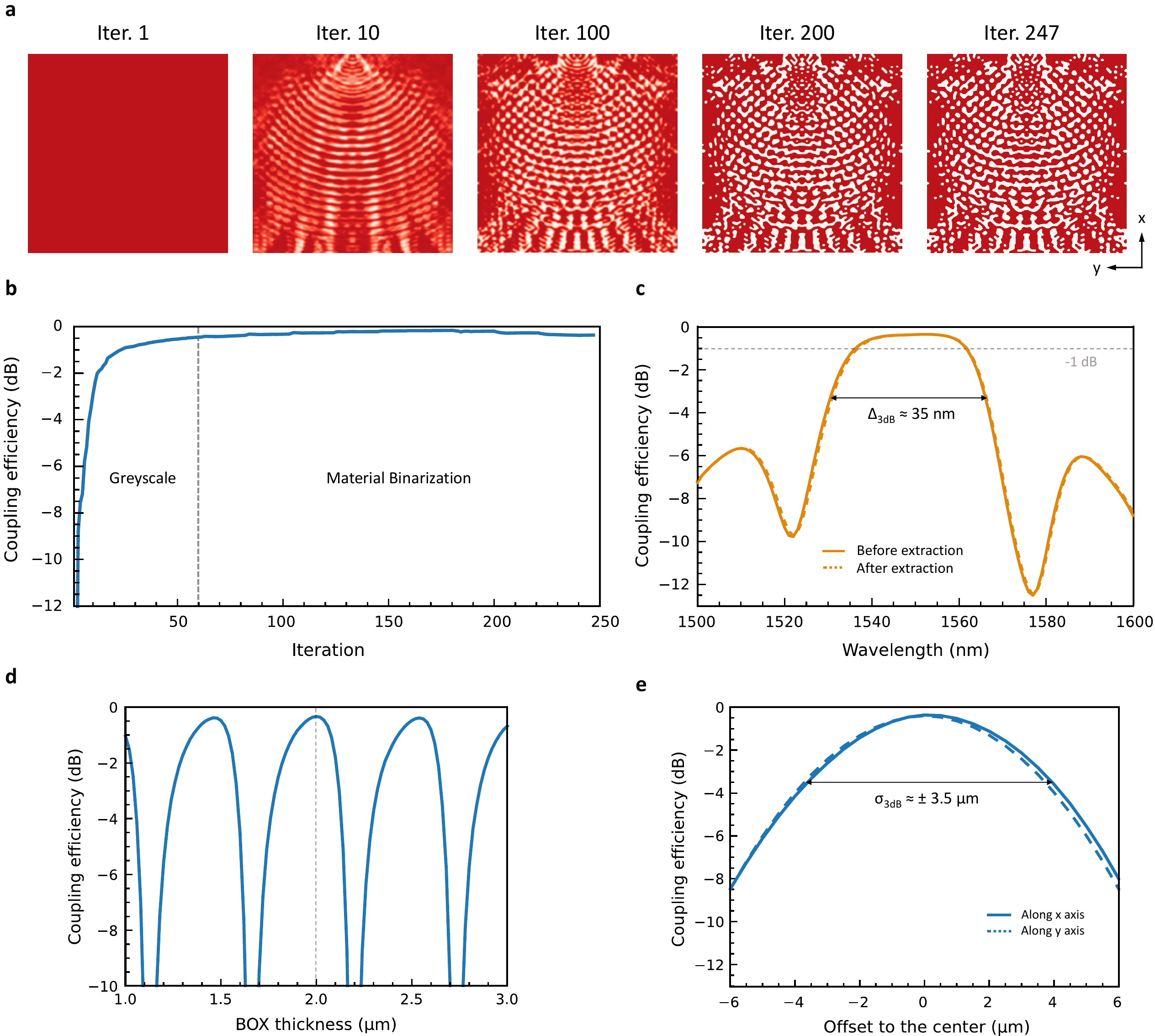}
\caption{
    (a) The snapshots of the topology-optimized structure in iterations 1, 10 100, 200 and 247. The refractive index of the material in the design region is presented in a gradient color scheme, 
    where the maroon (white) color corresponds to the reflective index of silicon (SiO$_2$) and the colors in between represent that of the pseudo mixture of the two materials.
    (b) Evolution of the coupling efficiency at the wavelength of 1550 nm during the optimization.
    (c) Simulated coupling efficiency of the topology-optimized vertical coupler with respect to the wavelength before (solid line) and after extraction (dashed line).
    The 3-dB bandwidth $\Delta_{\mathrm{3dB}} \approx$ 35 nm.
    (d) Coupling efficiency of the acquired topology-optimized vertical coupler at 1550 nm versus the thickness of the BOX layer.
    The dashed line indicates the coupling efficiency when the BOX thickness is 2 \textmu m.
    (e) Coupling efficiency of the acquired topology-optimized vertical coupler at 1550 nm versus the position offset along the x (solid line) and y (dashed line) axes.
    The alignment error giving rise to an additional 3-dB loss $\sigma_{\mathrm{3dB}}$ is around $\pm$ 3.5 \textmu m along the x and y axes from the center of the coupler.
    }
\label{fig:evolution and structures}
\end{figure}

The topology optimization starts with "bulk initial condition", i.e. initially the design region is full of silicon (iteration 1 in Fig.~\ref{fig:evolution and structures}(a)).
At the beginning of the optimization, the distribution of the reflective index varies in a greyscale fashion and first becomes like focusing periodic gratings (iteration 10 in Fig.~\ref{fig:evolution and structures}(a)).
Following multiple iterations, it is transformed into slender hollows and islands (iteration 100 in Fig.~\ref{fig:evolution and structures}(a)). 
The device performance also grows quickly to -2 dB within 10 iterations (Fig.~\ref{fig:evolution and structures}(b)).
After 60 iterations, the optimization enters the phase of material binarization where the material in each pixel in the design region gradually becomes either silicon or SiO$_2$. 
The device performance increases further but later drops slightly. It eventually converges at -0.35 dB (92.2\%).
The total iteration number of the optimization is 247 and the duration of the optimization is approximately 10 days using a desktop equipped with Intel i9-13900K and DDR5 random-access memory.
After the optimization is completed, the final structure is simulated with the FDTD method subsequently with a broader wavelength range and higher spectral resolution, namely from 1500~nm to 1600~nm with an increment of 1~nm, 
to ensure the acquired vertical coupler performs decently in the desired bandwidth.
The simulated coupling efficiency of the vertical coupler shows a flat response within the telecom C band (solid orange line in Fig.~\ref{fig:evolution and structures}(c)).
At the wavelength of 1550~nm, the coupling efficiency is consistently -0.35~dB with a 3-dB bandwidth $\Delta_{\mathrm{3dB}}$ of 35~nm.
Particularly, the bottom reflector recycles the direct transmission towards the substrate and thus considerably enhances the coupling efficiency.
The importance of the bottom reflector can also be stressed by the notable increase in the substrate leakage upon the exclusion of the reflector (grey line in Fig.~S1 in Supporting Information).

Afterward, the structure of the coupler is extracted, converted into chip layout files of standard format (the graphic data stream file in this case) and imported into the three-dimensional FDTD simulation environment.
We then repeat the FDTD simulation to ensure the integrity of the coupler's performance after the extraction.
The simulation result shows that the performance of the topology-optimized vertical coupler remains almost unaffected, as the spectral response shifts to the right marginally (dashed orange lines in Fig.~\ref{fig:evolution and structures} (c)).
In addition, since the directionality of an out-of-plane coupler is usually sensitive to the distance between the silicon device layer and the silicon substrate, 
we sweep the BOX thickness to ensure it is optimal for constructive reflection towards the coupler structure (Fig.~\ref{fig:evolution and structures}(d)).
Due to the phenomenon of wave interference, the coupling efficiency of the coupler at 1550 nm is periodic (having a period of $\sim$ 520 nm) with respect to the BOX thickness.
Besides, the coupling efficiency is maximized when BOX thickness is 2 \textmu m, which is a predetermined parameter in the optimization.
This indicates that, given a certain BOX thickness, structures that facilitate constructive interference after the light is reflected by the bottom reflector are explored during the topology optimization.
Finally, we investigate the alignment tolerance of the acquired vertical coupler along x and y directions (solid and dashed lines in Fig.~\ref{fig:evolution and structures}(e), respectively).
Only the coupling efficiency with respect to the offset along the y-axis is symmetric due to the x-axis symmetry of the acquired design (referred to the result of iteration 247 in Fig.~\ref{fig:evolution and structures}(a) and the rendered image in Fig.~\ref{fig:discussion}(a)).
The lateral alignment error resulting in an additional 3-dB loss $\sigma_{\mathrm{3dB}}$ is around $\pm$ 3.5 \textmu m from the center of the coupler along the x and y axes.
In this sense, common alignment routines with, for instance, piezo positioning stages should be applicable to this topology-optimized vertical coupler.

\clearpage

\clearpage

\section*{Discussion}
\begin{figure}[b!]
  \centering
  \includegraphics[scale=0.35]{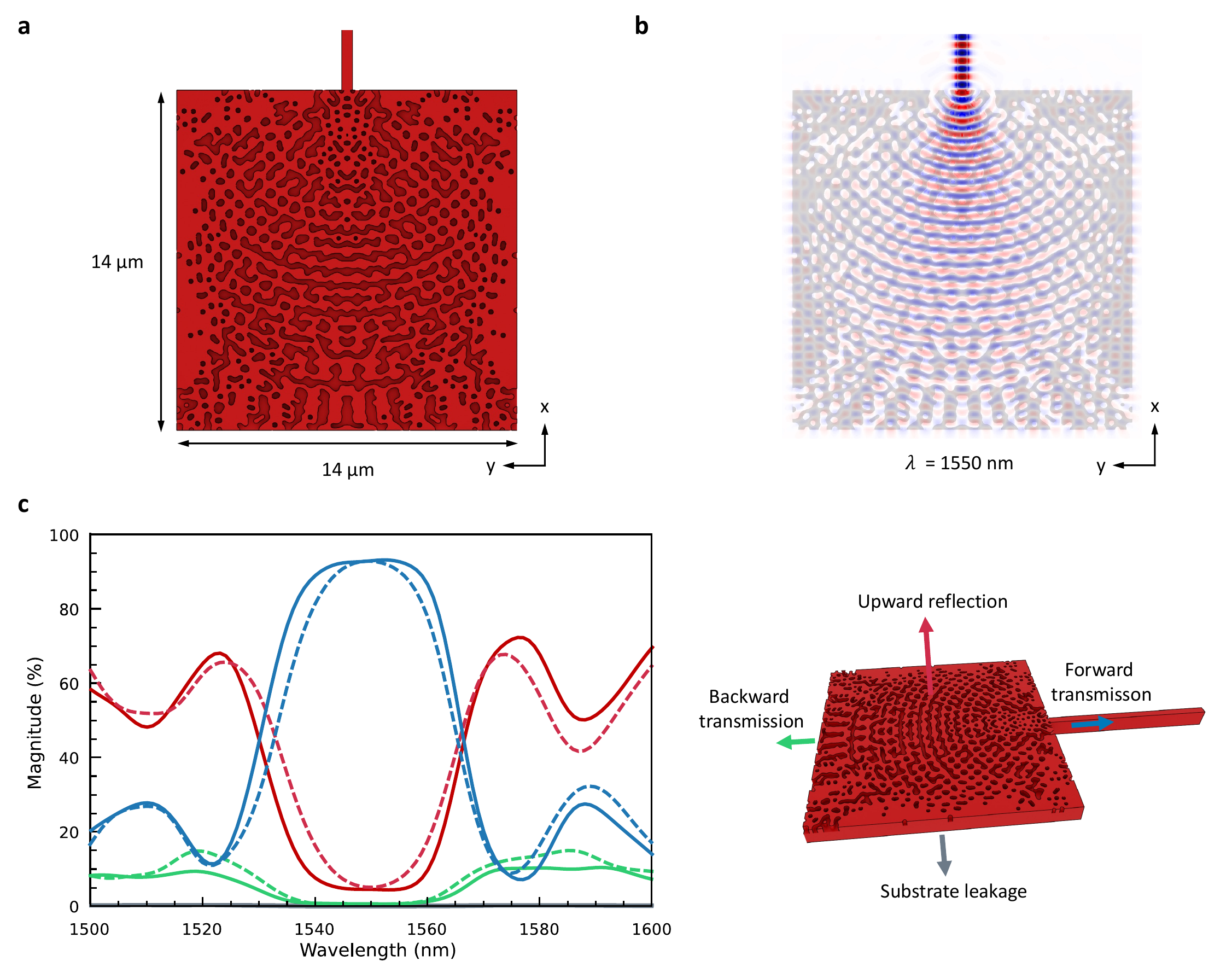}
  \caption{
  (a) Three-dimensional rendered image of the final design of the vertical coupler. The hollows in the rendered images should be filled with the cladding material, SiO$_2$, which is omitted in the image.
  (b) Simulated y-component of the electric field $E_y(x,y)$ at the wavelength of 1550~nm overlapped with the geometry of the topology-optimized vertical coupler. The input light in the simulation is y-polarized.
  (c) Left panel: forward (blue) and backward (green) transmission, upward reflection (red) and substrate leakage (grey) with respect to the wavelength. 
  The solid and dashed lines specify the optical responses of the designs acquired in two scenarios, namely (i) the optimization including the bottom reflector and (ii) reintroducing the bottom reflector after the optimization without the bottom reflector, respectively.
  Right panel: schematic for the forward and backward transmission, upward reflection and substrate leakage.
  }
  \label{fig:discussion}
\end{figure}

The optimization at the end leads to a "hole-based" structure as the algorithm seemingly focuses on forming hollows on the bulk silicon to diffract and couple the light into the strip waveguide efficiently (Fig.~\ref{fig:discussion}(a)).
Overall, the final design contains slender hollows and islands and nearly circular or elliptical holes.
The width of the long chains of islands and hollows and the diameter of the holes are roughly equal to or larger than 150 nm, which fits the constraint we preset on the minimum feature size in the optimization.
One can vaguely observe certain periodic patterns in the whole structure.
Such patterns are assumed capable of modulating the grating diffraction mode to improve the mode matching condition with the target Gaussian mode profile, akin to the conventional grating structure engineering methods like randomization \cite{zaoui2014bridging} or apodization \cite{marchetti2017high} of the grating period. 
Besides, the resulting design features an arc-like geometry over the middle region as the hollows and islands are arranged into confocal chains (Fig.~\ref{fig:discussion} (a)).
Similar to the focusing grating couplers \cite{van2007compact, wang2014focusing,hoppe2019ultra}, such a geometry is in principle advantageous to adiabatically transform the input optical mode into the fundamental mode of the strip waveguide while the whole structure only occupies a relatively small area.
This phenomenon can also be verified in the simulated distribution of the electric field component $E_y(x,y)$ at the wavelength of 1550 nm (Fig. \ref{fig:discussion}(b)).
After the light is coupled into the vertical coupler from the top, its mode profile becomes concavely shaped and then is gradually transformed into the fundamental mode of the strip waveguide.
In addition, the geometry near the bottom edge is assumed to function as a distributed Bragg reflector to redirect the light back to the strip waveguide, as the backward transmission of the structure is strongly suppressed within the range of 1540 nm to 1560 nm (solid green line in Fig.~\ref{fig:discussion}(c)).
Despite the high coupling efficiency, the major loss, which is the upward reflection of $\sim 4.3\%$ (solid red line in Fig.~\ref{fig:discussion}(c)), comes from the fact that there is still a certain amount of the light traveling through the topology-optimized structures after being reflected by the bottom reflector.
Nevertheless, owing to the bottom reflector, the substrate leakage is drastically reduced and is thus nearly zero (solid grey lines in Fig.~\ref{fig:discussion}(c)).
We also consider the scenario where the bottom reflector is excluded in the topology optimization and reintroduced after the final design is delivered (geometry of the resulting design and other details are shown in Fig.~S2 in Supporting Information).
However, we observe that the upward reflection increases within the range of 1530 nm to 1570 nm (dashed red line in Fig.~\ref{fig:discussion}(c)) after reinserting the bottom reflector underneath the 2-µm BOX layer.
This implies that the design generated in this manner is suboptimal for recoupling the reflected light within the preset wavelength range.
Despite the comparable peak performance, this results in a narrower operating bandwidth compared to the original design (dashed and solid blue lines in Fig.~\ref{fig:discussion}(c)).
Therefore, incorporating the bottom reflector into the topology optimization is critical for optimizing the device performance in the desired bandwidth.

It is also possible to have various designs of the topology-optimized vertical coupler by choosing other initial conditions to expand the component library.
For instance, the topology optimization can begin with the "mixture initial condition", i.e.~a pseudo material of which the refractive index is the average of silicon and SiO2 is in the discretized design region initially.
This gives rise to an "island-based" geometry (Fig.~S3 (a) in Supporting Information), as the algorithm seemingly tends to transform the pseudo material into long silicon islands that can guide the light into the strip waveguide.
According to our simulation results, both island- and hole-based topology-optimized vertical couplers have similar spectral responses and peak performances (Fig.~S3 (b) in Supporting Information).
One may then select a suitable design based on, for instance, the preference or the capability of a chip foundry.
There are several possible ways to further improve the design and the performance of the topology-optimized coupler.
For instance, the operating wavelength or the 3-dB bandwidth of the vertical coupler can be further expanded if a broader wavelength range in which the FoM function is maximized is used in the topology optimization.
In this case, however, more wavelength points should be considered in the optimization to obtain a flat and usable response of the devices, which is more computationally expensive and may significantly increase the duration of the optimization.
In addition, initial conditions with various predefined geometries, e.g.~a focusing silicon grating structure with Bragg reflectors put close to the back edge, in the discretized design region could also be introduced to explore new designs of vertical on-chip couplers.
Ultimately, one may modify the optimization algorithm that also parameterizes the depth of the design region to produce a topology-optimized structure that contains full-etched and shallow-etched hollows.
Effectively, this type of topology-optimized coupler could behave like the interleaved \cite{alonso2014fiber,benedikovic2015high} or L-shaped grating couplers \cite{benedikovic2017shaped,watanabe2017perpendicular} as the dual-etched slots strengthen the diffraction effect and improve the directionality of light propagation.
In this way, the coupler could suppress the loss caused by the light that travels toward the silicon substrate even without a bottom reflector, which greatly reduces the cost and complexity of the fabrication.

\section*{Conclusion}
In summary, we inversely design a compact vertical coupler on the 220-nm SOI platform via topology optimizations and provide an in-depth investigation of the outcome.
The predicted coupling efficiency of the topology-optimized vertical coupler, which incorporates a bottom reflector, is -0.35 dB at the wavelength of 1550 nm and the 3-dB bandwidth is 35 nm.
As the area of the couplers is only 14 \textmu m $\times$ 14 \textmu m, the footprint in a chip layout can be reduced considerably compared to conventional grating couplers.
In addition, the dimensions of the sophisticated features in the design, such as the diameter of the small holes and the width of the long chains of slender islands and hollows, are in general equal to or greater than 150 nm,
which ensures the fabricability on the SOI platform.
Altogether, we believe our study will facilitate practical applications of topology-optimized components, especially the couplers, further in various fields in integrated photonics.

\section*{Methods}
\subsection*{Numerical simulations and optimizations}
We exploit the inverse design tool, Lumopt, together with the finite-difference time-domain (FDTD) solver (Ansys Lumerical FDTD) to perform the topology optimization.
In the topology optimization, a design region with a thickness of 70 nm is set up on top of a silicon layer that has a thickness of 150 nm (Fig.\ref{fig:sim_sketch} (a)).
Both the design region and the silicon layer are connected to a strip waveguide that has the dimensions of 450 nm $\times$ 220 nm and the entire device is embedded in the SiO$_2$ cladding material (Fig.\ref{fig:sim_sketch} (b)).
The space in the design region is discretized into pixels parameterized by $\rho(x,y) \in [0, 1]$, where $x$ and $y$ denote the position of the pixel in the x and y directions.
In each pixel, the permittivity of the material $\epsilon$ is in the linear interpolation
\begin{eqnarray}
  \epsilon(x,y) =  \epsilon_{\rm{SiO_2}} + \rho(x,y) \cdot (\epsilon_{\rm{Si}}-\epsilon_{\rm{SiO_2}})
\end{eqnarray}  
and then adjusted iteratively to maximize user-defined FoM functions by solving the optimization problem.
Eventually, the material in each pixel becomes either solid (silicon) or void (SiO$_2$) by applying the Heaviside function
\begin{eqnarray}
  \tilde\rho(\rho) = \frac{\tanh(\beta \eta)+ \tanh[\beta (\rho-\eta)]}{\tanh(\beta \eta)+ \tanh[\beta (1-\eta)]}  
\end{eqnarray}
for threshold projection, where $\beta$ and $\eta$ are the threshold parameters.
In this material binarization phase, the permittivity of the material in each pixel, given by
\begin{eqnarray}
\epsilon(x,y) =  \epsilon_{\rm{SiO_2}} + \tilde\rho(x,y) \cdot (\epsilon_{\rm{Si}}-\epsilon_{\rm{SiO_2}}),
\end{eqnarray}
becomes binary after ramping up $\beta$ to form a physically-feasible structure. 

The size of the design region is configured to be 14 \textmu m $\times$ 14 \textmu m $\times$ 70 nm.
The 14-\textmu m length and width are set up for a better accommodation of the Gaussian profile of the light from the single-mode fiber, considering a mean field diameter of an SMF-28 fiber of around 10.4 \textmu m.
The 70-nm thickness is planned for the creation of shallow-etched sub-wavelength topology-optimized structure that is able to diffract the light efficiently into the strip waveguide.
The light source in the simulations is set to be a Gaussian source positioned at the center of the design region. 
Its polarization state is perpendicular to the strip silicon waveguide (i.e.~y-polarized) and the mode field diameter is set to be 10.4 \textmu m, mimicking the light emitted from an SMF 28 optical fiber.
The incident angle of the Gaussian light source is also set to 0\textdegree \  to fulfill the vertical fiber-coupling scheme.
A perfect electric conductor layer is then placed underneath a 2-\textmu m buried oxide layer and serves as a metal bottom reflector to reflect the light that propagates through the coupler (Fig.\ref{fig:sim_sketch} (b)) and let the coupler recouple the light into the waveguide.
Such a reflector is commercially available and can be implemented by depositing the metal layer in the trenches etched on the backside of the wafer and beneath the couplers \cite{zaoui2014bridging} or using the benzocyclobutene (BCB) bonding method where the metal layer expands all over the wafer \cite{higuera2015realization,bleiker2017adhesive}.

To obtain a flat and continuous spectral response in the operating wavelength, a set of 5 wavelength points, from 1540 nm to 1560 nm with an increment of 5 nm, at which the FoM function is maximized is preset in the optimization.
Since the final setup has a mirror symmetry along the x direction, the boundary conditions of the FDTD simulations is set to be anti-symmetric at the y-boundary in order to reduce the computation time by a factor of 2 in each FDTD simulation. 
Essentially, the FoM function in the topology optimization is defined as the overlap between the input mode in the strip waveguide and the fundamental mode of the strip waveguide, which indicates the efficiency of the light coupling into the waveguide fundamental mode.
Finally, a constraint on the minimum feature size of the topology-optimized structure is set to 150 nm to guarantee the fabricability of the final structures.

During the topology optimization, three-dimensional FDTD simulations are initialized and carried out with the settings mentioned above.
Such a single FDTD simulation in general requires around 30 minutes and the gradient of the FoM function in the parameter space is calculated based on the simulation results.
Within the design region, the refractive index of the material in the pixels is then updated according to the gradient of the FoM function computed previously for the next iteration to maximize the FoM function.
Owing to the adjoint method applied in the topology optimization \cite{lalau2013adjoint}, in general, only two FDTD simulations, commonly named forward and adjoint simulations, are required to determine the gradient of the FoM function, which greatly improves the efficiency of the whole optimization.
Still, since the adjoint method is a gradient-descent method, the initial condition of the design region may significantly influence the final output of the topology optimization.

\begin{acknowledgement}
The research was funded by Bundesministerium für Bildung und Forschung (PhotonQ, SiSiQ), Carl-Zeiss-Stiftung, Competence Center Quantum Computing Baden-Württemberg (Project QORA) and Deutsche Forschungsgemeinschaft (431314977, GRK2642).
\end{acknowledgement}

\bibliography{sample}

\end{document}


\newpage

\section{Simulation of the final design after removing the bottom reflector}
\begin{figure*}[ht]
  \centering
  \includegraphics[scale=0.42]{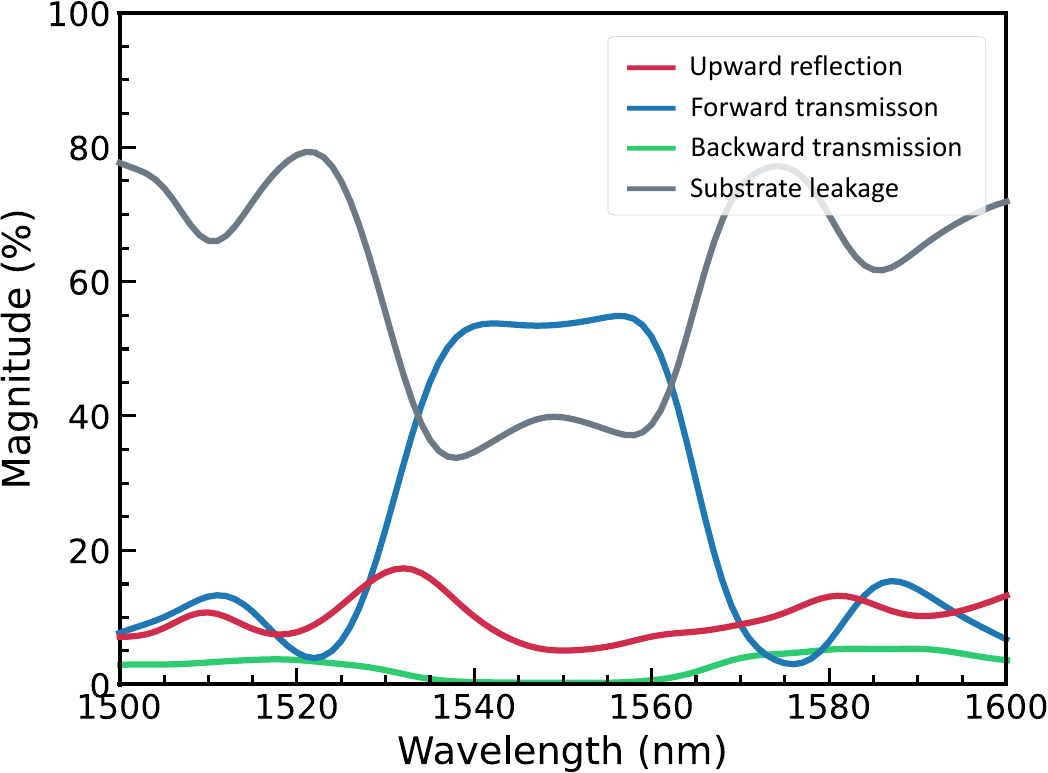}
  \caption{
    Forward (blue) and backward (green) transmission, upward reflection (red) and substrate leakage (grey) with respect to the wavelength after removing the bottom reflector from the final design. 
  }
  \label{fig:no_mirror_results}
  \end{figure*}
After we acquire the design of the vertical coupler produced using topology optimization incorporating the bottom reflector, the reflector is removed from the coupler to study the effect it exerts.
In the three-dimensional FDTD simulation,
the Gaussian light source with y-polarization is positioned at the center of the coupler and injected from the top.
The thickness of the buried oxide layer is still set to be 2 µm to have a fair comparison with the original design with a bottom reflector.
The simulation is then performed within the wavelength range of 1500 nm to 1600 nm with a resolution of 1 nm.

After omitting the bottom reflector, the substrate leakage increases drastically to around 40\% in the range of 1530 nm to 1560 nm (solid grey line in Fig.~\ref{fig:no_mirror_results}) while the upward reflection back to the optical fiber and the backward transmission remain relatively low (solid red and green lines in Fig.~\ref{fig:no_mirror_results}, respectively).
Due to the massive loss towards the substrate, the forward transmission that correlates to the coupling efficiency of the coupler is reduced to around 55\% at 1550 nm (solid blue line in Fig.~\ref{fig:no_mirror_results}).
Therefore, the bottom reflector is important for improving the overall coupling efficiency since it enhances the directionality of the topology-optimized vertical coupler.

\newpage

\section{Exclude the bottom reflector in the topology optimization}
\begin{figure}[h]
  \centering
  \includegraphics[scale=0.42]{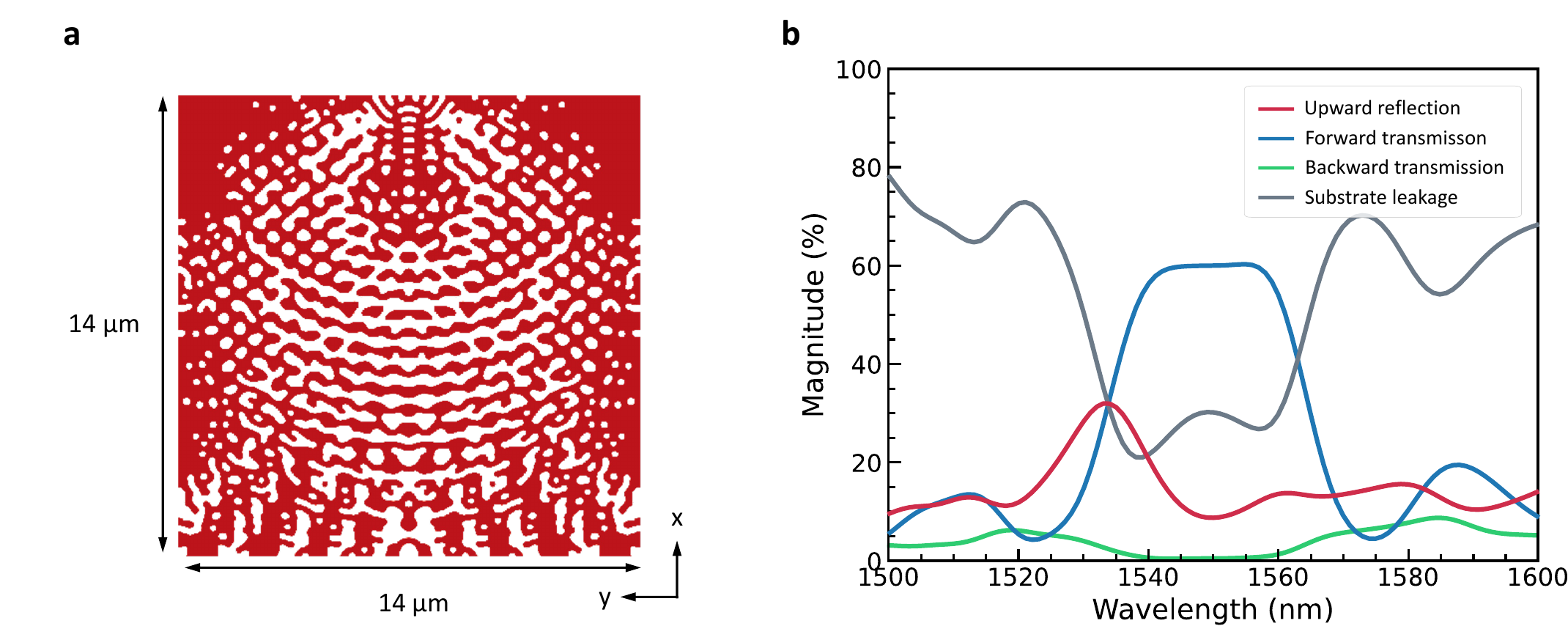}
  \caption{
    (a) The refractive index distribution of the design delivered by the optimization excluding the bottom reflector. 
    The maroon (white) color corresponds to the reflective index of silicon (SiO$_2$).
    (b) Forward (blue) and backward (green) transmission, upward reflection (red) and substrate leakage (grey) with respect to the wavelength of the design produced by the optimization excluding the bottom mirror.
    }
  \label{fig:no mirror results}
  \end{figure}
We also apply the topology optimization where the bottom reflector is excluded in the design of a vertical coupler. 
The settings regarding the simulation and optimization remain the same as the descriptions outlined in Method section in the main text.
The total iteration number of the optimization in this case is 393 and the duration of the optimization is roughly 12 days.
The resulting design (Fig.~\ref{fig:no mirror results}(a)) shares a similar feature with the design delivered by the optimization incorporating a bottom reflector. 
The simulated coupling efficiency is -2.23 dB (59.8\%) at the wavelength of 1550 nm.

\newpage

\section{Topology-optimized vertical coupler delivered using a different initial condition}
\begin{figure}[h!]
  \centering
  \includegraphics[scale=0.42]{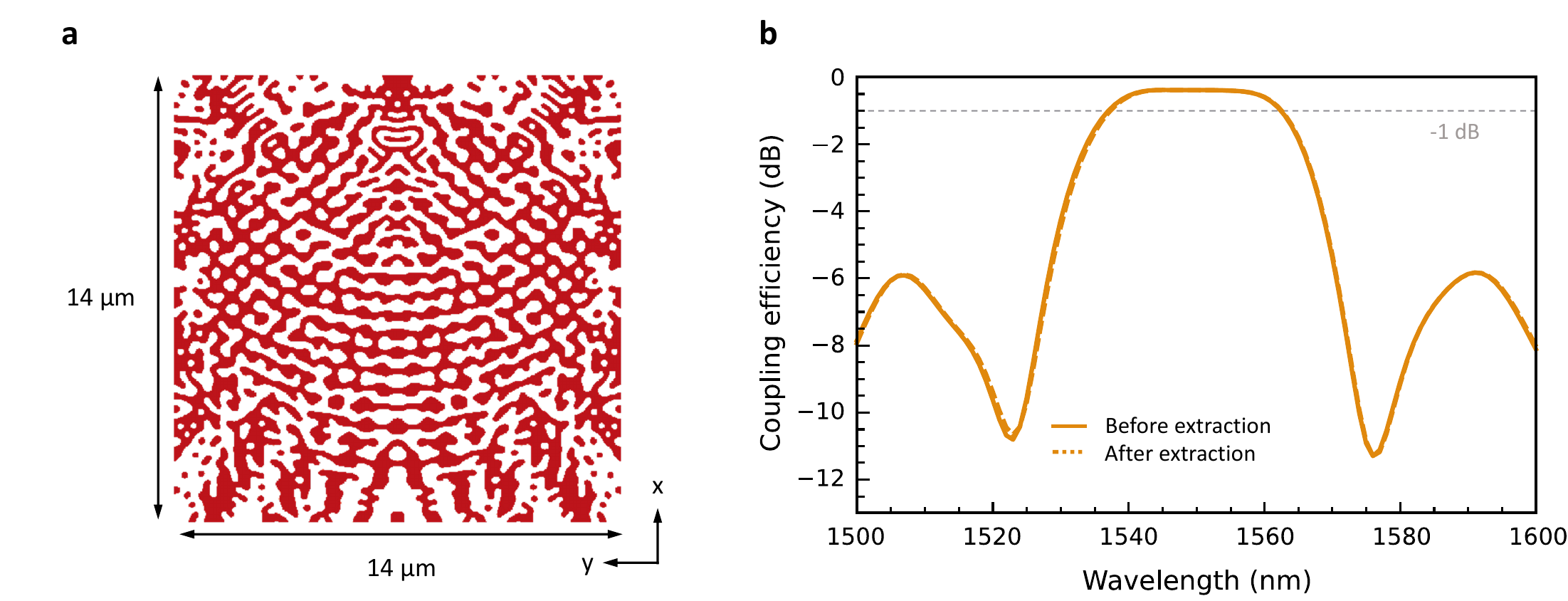}
  \caption{
        (a) The refractive index distribution of the design delivered by the optimization beginning with the mixture initial condition. 
    The maroon (white) color corresponds to the reflective index of silicon (SiO$_2$).
    (b) Simulated coupling efficiency of the island-based topology-optimized vertical coupler with respect to the wavelength before (solid line) and after extraction (dashed line).
    }
  \label{fig:island-based results}
  \end{figure}

As the adjoint method is a gradient-descent method, the initial condition of the design region may significantly influence the final output of the topology optimization.
Here we demonstrate the result of the topology optimization using the "mixture" initial condition.
In this case, a pseudo material of which the refractive index is the average of silicon and SiO2 is set in the design region initially.
Other settings regarding the simulation and optimization remain the same as the descriptions outlined in the Method section in the main text.
After the optimization is completed, the final structure (Fig.~\ref{fig:island-based results} (a)) is also simulated subsequently with a broader wavelength range and higher spectrum resolution.

The total iteration number of the optimization is 246, which takes around 10 days. The simulated coupling efficiency of the "island-based" coupler across the wavelength from 1500 nm to 1600 nm shows comparable performance with the "hole-based" coupler shown in the main text.
At the wavelength of 1550 nm, the coupling efficiency is -0.374 dB (91.7\%).
The structure of the coupler is also extracted and imported into the FDTD simulation environment for another subsequent simulation.
The simulation result shows that the performance of the device remains almost unaffected (Fig.~\ref{fig:island-based results} (b)).